\documentclass{aa}  

\usepackage{graphicx}
\usepackage{txfonts}
\newcommand{\hii}{\relax \ifmmode {\mbox H\,{\scshape ii}}\else H\,{\scshape ii}\fi}
\newcommand{\mi}{\relax \ifmmode {\mu{\mbox m}}\else $\mu$m\fi}
\newcommand{\ha}{\relax \ifmmode {\mbox H}\alpha\else H$\alpha$\fi}
\newcommand{\hb}{\relax \ifmmode {\mbox H}\beta\else H$\beta$\fi}

\newcommand{\sii}{\relax \ifmmode {\mbox S\,{\scshape ii}}\else S\,{\scshape ii}\fi}
\newcommand{\siii}{\relax \ifmmode {\mbox S\,{\scshape iii}}\else S\,{\scshape iii}\fi}
\newcommand{\nii}{\relax \ifmmode {\mbox N\,{\scshape ii}}\else N\,{\scshape ii}\fi}
\newcommand{\oi}{\relax \ifmmode {\mbox O\,{\scshape i}}\else O\,{\scshape i}\fi}
\newcommand{\oii}{\relax \ifmmode {\mbox O\,{\scshape ii}}\else O\,{\scshape ii}\fi}
\newcommand{\hei}{\relax \ifmmode {\mbox He\,{\scshape i}}\else He\,{\scshape i}\fi}
\newcommand{\heii}{\relax \ifmmode {\mbox He\,{\scshape ii}}\else He\,{\scshape ii}\fi}
\newcommand{\oiii}{\relax \ifmmode {\mbox O\,{\scshape iii}}\else O\,{\scshape iii}\fi}
\newcommand{\neiii}{\relax \ifmmode {\mbox Ne\,{\scshape iii}}\else Ne\,{\scshape iii}\fi}

\newcommand{\rdostres}{\relax \ifmmode {\,\mbox{R}}_{\rm 23}\else \,\mbox{R}$_{\rm 23}$\fi} 

\newcommand{\ciii}{\relax \ifmmode {\mbox O\,{\scshape iii}}\else C\,{\scshape iii}\fi}
\newcommand{\civ}{\relax \ifmmode {\mbox O\,{\scshape iii}}\else C\,{\scshape iv}\fi}

\newcommand{\gsim}{\hbox{\rlap{\lower.55ex\hbox{$\sim$}} \kern-.3em
\raise.4ex \hbox{$>$}}}
\newcommand{\lsim}{\hbox{\rlap{\lower.55ex\hbox{$\sim$}} \kern-.3em
\raise.4ex \hbox{$<$}}}

\usepackage[normalem]{ulem}
\usepackage{color}

\begin{document}

   \title{Photon leaking or very hard ionizing radiation? Unveiling the nature of \heii -emitters using the softness diagram}

\titlerunning{\heii -emitters from the softness diagram}

   \author{E. P\'erez-Montero
          \inst{1}
          \and
          C. Kehrig\inst{1}
          \and
          J. M. V\'\i lchez\inst{1}
          \and
          R. Garc\'\i a-Benito\inst{1}
          \and
          S. Duarte Puertas\inst{1}
          \and
          J. Iglesias-P\'aramo\inst{1}
 }

   \institute{Instituto de Astrof\'\i sica de Andaluc\'\i a - CSIC. Apdo. 3004. E-18080. Granada. Spain\\
              \email{epm@iaa.es}
             }

   \date{Received XXX; accepted YYY}

 
  \abstract
   {} 
 {Star-forming galaxies with nebular \heii\ emission contain very energetic ionizing sources of radiation, which can be considered as analogs to the major contributors of the reionization of the Universe in early epochs.  It is therefore of great importance to provide a reliable absolute scale for the equivalent effective temperature ($T_*$) for these sources.} 
{We  study a sample of local ($z <$ 0.2) star-forming galaxies showing optical nebular \heii\ emission using the so-called softness diagrams, involving emission lines of two elements in two consecutive stages of ionization (e.g., [\sii]/[\siii] vs. [\oii]/[\oiii]).  We use for the first time the \hei/\heii\ ratio in these diagrams in order to explore the higher range of $T_*$ expected in these objects, and to investigate the role of possible mechanisms driving the distribution of galaxy points in these diagrams. We build grids of photoionization models covering different black-body temperatures, model cluster atmospheres, and density-bounded geometries to explain the conditions observed in the sample.}
{We verified that the use of the softness diagrams including the
  emission-line ratio \hei/\heii\ combined with black-body photoionization models can provide an absolute scale of
          $T_*$ for these objects. The application of a Bayesian-like
          code indicates $T_*$ in the range 50-80 kK for the sample of
          galaxies, with a mean value higher than 60 kK. The average
          of these high temperature values can only be reproduced
          using  cluster model populations with nearly metal-free
          stars, although such ionizing sources cannot explain either
          the highest $T_*$ values, beyond 1$\cdot \sigma$, or the
          dispersion observed in the softness diagrams.  According to our photoionization models, most sample galaxies could be
affected to some extent by ionizing photon leaking, presenting a mean
photon absorption fraction of 26\% or higher depending on the
metallicity assumed for the ionizing cluster. The entire range of
\hei/\heii\  , [\sii]/[\siii], and [\oii]/[\oiii] ratios for these HeII-emitting galaxies is reproduced with our models, combining nearly metal-free ionizing clusters and photon leaking under different density-bounded conditions.}
  {}


   \maketitle
%

\section{Introduction}

The presence of \heii\ recombination emission lines (e.g., at
$\lambda$1640 \AA\ and $\lambda$4686 \AA\ in the rest-frame UV and
optical ranges, respectively) in the spectra of star-forming (SF)
galaxies is evidence of very hard and energetic sources of ionizing
radiation with energy beyond 4 Ryd. High-ionization lines, such as
\heii\,
are found to be more frequent in high-redshift galaxies than locally, and
SF galaxies with lower metal content tend to have larger nebular \heii\
line intensities than those with higher metallicities \citep[e.g.,][]{guseva2000,kehrig11,shirazi12,cassata13,nanay19,saxena19}. This
agrees with the expected harder spectral energy distribution (SED) at
the lower metallicities typical in the far-away Universe \citep[e.g.,][]{smith15,st19}.
The interpretation and understanding  of the presence of
these high-excitation lines in SF objects in the young Universe could therefore  be of great relevance for the preparation of upcoming extragalactic surveys such as EUCLID \citep{euclid}, JADES \citep{jades}, or WFIRST (renamed as NGRST, \citealt{akeson19}). 

Theoretical arguments suggest that PopIII stars and nearly metal-free ($Z$ $<$
$Z_{\odot}$/100) stars have spectra that are hard enough to produce many
He$^{+}$-ionizing photons, and so the high-ionization HeII line has
been considered one of the most useful signatures to single out
candidates for the elusive PopIII-hosting galaxies \citep[e.g.,][]{ts00,yoon12,v15}.
Hot massive stars, shocks, and X-ray binaries are among the most
popular candidates for producing nebular \heii\ emission in local SF
objects \citep[e.g.,][]{garnett91,cer02,ti05,shirazi12,k15,s20}. However,
despite intense observational and theoretical efforts over recent
years, \heii\ ionization is still puzzling, especially in low-metallicity
galaxies \citep[e.g.,][]{garnett91,bpass,k18,gotberg18,st19,ku19,plat19,s20}.
Recently, \cite{k15,k18} studied  the spatial distribution of
nebular HeII emission in detail in the two extremely metal-poor galaxies (XMPs; $Z < Z_{\odot}$/10) IZw18 and SBS0335-052E, and find that only hot massive stars with
metallicity much lower than that of their HII regions can explain the
observations.

One of  the main reasons for our lack of understanding of the physics behind the
nebular \heii\ emission is  our inability to obtain direct observations of the ionizing continuum of massive
stars at $\lambda$ $<$ 228~\AA~ (He$^{+}$ ionization edge) at any
redshift    with current (modern) facilities. Obtaining a detailed comprehension
of such stellar SEDs, which are notoriously difficult to model and
suffer from several uncertainties \citep[e.g.,][]{crowther02,kubat12,dori15,ku19}, continues to be a major challenge. One of the goals of this work is
to constrain the little-known SEDs of metal-poor, hot massive stars,
which is very important in our quest to better understand the HeII-emitting gas
properties in low-metallicity environments.

Among the different tools to ascertain the nature of the ionizing
stellar clusters in gaseous nebulae, which relies on
the availability of the most prominent collisional optical emission lines,
is the radiation softness parameter ($\eta$), defined by \cite{vp88}. The parameter $\eta$ can be used to obtain a scale of the
hardening of the ionizing radiation field, and it can be accessed from the optical spectrum
with no need for the UV continuum.
As shown below, $\eta$ is based on the relative ratios of the
abundances of two different species in two consecutive stages of ionization, such as for example

\begin{equation}
\eta = \frac{O^+/O^{2+}}{S^+/S^{2+}}
\end{equation}

In objects where it is difficult to precisely measure the ionic abundances,  an alternative formulation can be used which is based on emission-line flux ratios; for example, using the following emission lines,

\begin{equation}
\eta\prime = \frac{[OII] \lambda 3727/[OIII] \lambda\lambda 4959,5007}{[SII] \lambda\lambda 6717,6731 / [SIII] \lambda\lambda 9069,9532}
\end{equation}

Both expressions for the softness parameter decrease for higher values of  $T_*$  \citep[][]{vp88}, though a certain dependence on metallicity \citep{morisset04} and on the ionization parameter ($U$)\citep{pmv09} has to be considered. These dependencies can be minimized using the emission-line ratios involved in the $\eta$ parameter in a diagram, which we denominate here as the softness or $\eta$ diagram (e.g., [\sii]/[\siii] vs. [\oii]/[\oiii]), and so models for different $T_*$ can be compared directly with observations on this plane
\citep{pm14,fm17}.

In this work we study the $\eta$ diagram, introducing the ionic ratio of nebular He lines (i.e., \hei/ \heii) in order to
better trace the slope of the ionizing SED at the frequencies responsible for the ionization of He$^+$.
The ratio between the flux of \heii\ $\lambda$ 4686 and the continuum flux at certain bands  
emitted from the gas ionized by a central ionizing star has already been used as an indicator of its $T_*$
such as in planetary nebulae (PNe)
following the so-called Zanstra method \citep[e.g.,][]{phillips04}.
 In addition, the relation between the emission-line ratio of \hei\ $\lambda$ 5876 \AA\ to \heii\ $\lambda$ 4686 \AA\ with $T_*$ in low-density astrophysical plasmas is very well established \citep{Smits1996} and has been used in hot PNe
\citep[e.g.,][]{Seaton1960,Ratag1997}, and in the gas diagnostics in
the broad-line region of active galactic nuclei \citep[(AGNs); e.g.,][]{corista2004,dragana2010}.  
Therefore, this emission-line ratio
 can also be used in this context in combination with other ratios of emission lines of
consecutive ions to reduce the dependence on $U$ and to provide a scale of $T_*$ in
SF galaxies with \heii\ emission.

On the other hand, certain emission-line ratios of consecutive ions (e.g., the relation of the  O32 = [\oiii] $\lambda$ 5007/[\oii] $\lambda$ 3727 ratio versus the R23-index) in SF galaxies have been used to estimate the fraction of escaping photons \citep[e.g.,][]{nakajima14,Paalvast2018}.
In addition, \heii\ $\lambda$ 4686 \AA\  is sensitive to  both
mass-loss rates (via stellar winds) and $T_*$ \citep{Massey2013}. Recently, \cite{izotov16} and \cite{izotov18}  quantified the escape fraction of Lyman
continuum (LyC) ionizing photons in a sample of SF galaxies; a few of these LyC leakers show nebular \heii\ emission \citep[e.g.,][]{schaerer18}. 
 This implies that the presence of a very hard ionizing spectrum is not a 
required condition for LyC emission. Nonetheless, we cannot exclude that a
significant fraction of LyC photons can escape from some nebular
HeII emitters.
Therefore, leaking is expected to substantially affect the different versions of the softness diagram.

The paper is organized as follows: in Section 2 we describe our
control sample of \heii \ emitters
 taken from \cite{shirazi12}, and how we reanalyzed their spectra and derived their main physical properties
and chemical abundances. In Section 3.1 we present the results of the behavior of this sample in the defined
$\eta$ diagrams. In Section 3.2 we introduce an adapted version of the model-based code {\sc Hii-Chi-mistry-Teff} (hereinafter {\sc HCm-Teff}, \citealt{pm19}) and we provide a $T_*$ scale for the studied \heii \ emitters. In Section 3.3 we discuss how stellar cluster
atmospheres at different metallicities can account for the derived $T_*$ scale and in Section 3.4 we explore how photon leaking can also be
invoked as an alternative explanation for the observed dispersion in the studied diagrams. In Section 4 we summarize our
results and conclusions. In Appendix A.1 we make use of the different sets of models to provide new expressions for the ionization correction factors (ICFs) used to derive the total oxygen abundance when the \heii\ emission line is detected.

\section{Control sample}

We used the list of nebular \heii \ emitters selected by   \cite{shirazi12}
from the seventh data release of the Sloan Digital Sky Survey (SDSS, \citealt{sdss-dr7}) and classified as SF galaxies. 
These SF \heii \ emitters are mainly local (0.001 $< z <$ 0.196) and
and some of them are galaxies
for which the optical broad emission at 4650 \AA\ has been detected, which is attributed to Wolf-Rayet (WR) stars if
adopting the classification suggested by \cite{shirazi12}.
In any case, given the heterogeneous redshift distribution of the sample and the different angular area covered
by the 3''-diameter SDSS fiber, this detection must be taken with caution in the lowest redshift objects; see for instance,
the impact of the aperture bias when integral field spectroscopy is used instead as discussed in \cite{Kehrig2013}, \cite{Miralles2016},
and \cite{Liang2020}  for more details.

We re-analyzed the nebular emission lines in the selected  spectra of the \heii \ emitters.
In a first step, we subtracted the underlying stellar population using the spectral synthesis code {\sc STARLIGHT} 
\citep{cid2004,cid2005}. {\sc STARLIGHT} fits an observed continuum SED using a combination 
of the synthesis spectra of different single stellar populations (SSPs). We chose the SSP spectra from \cite{bc2003} 
based on the {\sc STELIB} library from \cite{LeBorgne2003}, 
{\sc Padova} 1994 evolutionary tracks, and a \cite{Chabrier2003} initial mass function (IMF) between 0.1 and 100 M$_{\odot}$. 
Four metallicities were selected, from $Z$ = 0.0001 up to $Z$ = 0.008, with 41 ages spanning from 1 Myr up to 14 Gyr 
for each metallicity. The {\sc STARLIGHT} code builds a nonparametric linear combination of the different SSPs,  
simultaneously solving the ages, metallicities, and the average reddening. The reddening law from \cite{ccm89} 
with RV = 3.1 was used. Prior to the fitting procedure, the spectra
were shifted to the rest frame and corrected for Galactic extinction according to 
\cite{Schlegel1998}.

After subtracting the {\sc STARLIGHT} best-fitting  stellar model from the observed spectra, we analyzed the 
resulting residuals in the selected sample of \heii\ emitters using the {\sc SHIFU}\footnote{SHerpa IFU line fitting package (Garc\'\ a-Benito in
prep.).} package to obtain the flux of the emission lines. The package contains a suite of routines that can be used to easily analyze emission or absorption lines (both in cube and RSS format). 
Individual spectra, such as those employed in our case, can be provided as a list of RSS files. 
The core of the code uses the {\sc  CIAO} {\sc Sherpa} package \citep{freeman2001}. 
Several custom automatic algorithms are implemented in order to cope with general and ill-defined cases. 
Although the fit is performed in the stellar-continuum-subtracted spectra, we allowed for the modeling of the continuum to take into account small deviations 
in the stellar continuum residuals. 
A sigma clipping was independently applied to the residual spectra, and then this was parsed to the composite line plus (residual) continuum model. 
A first-order polynomial was chosen for the continuum, while single Gaussians were selected for the lines. 
The continuum was evaluated in the original spectra to determine equivalent widths. 
Uncertainties in the measured values are evaluated by perturbing the residual spectra according to the error vector 100 times.

Considering  only  line fluxes with a signal-to-noise ratio (S/N)
of greater than or equal to 3 for all our calculations in this work,
including the narrow nebular \heii\ $\lambda$4686 \AA\ emission line, we obtain a total of 194 objects, the same as those studied in \cite{shirazi12}.
However, for our analysis we ruled out 8 objects in the sample that   either have a relative \heii/\hb\ flux ratio larger than 0.1, as these can present aperture problems
at very low redshifts (e.g., Mrk~178 or SHOC~22), or present a  \heii\ line profile with a very conspicuous aspect after detailed visual inspection (e.g., NGC 4449).
This leaves a total of  186 objects in the final sample, which includes 106 objects that also present
the broad bump at around $\lambda$4650 \AA\ emitted by WR stars, according to \cite{shirazi12}.

All  emission-line fluxes were corrected for intrinsic extinction
using C(\hb) calculated comparing the measured and theoretical ratios
of all  Balmer hydrogen lines with S/N $\geq$ 3. We assume the case B
and the average physical conditions derived for our sample, and considered  the
extinction law from \cite{ccm89}.

Electron densities and temperatures were calculated using the 
[\sii] $\lambda$6730/$\lambda$6717 \AA\ and [\oiii] ($\lambda$4959+$\lambda$5007)/$\lambda$4363 \AA\ emission-line
ratios, respectively, and using the software {\sc pyneb} \citep{pyneb}. 
The measured mean and median values for the electron density are 100
particles per cm$^{-3}$ and 75 cm$^{-3}$, respectively. We were able to measure the
electron temperature for 167 objects 
directly  using [\oiii]$\lambda$4363 \AA, giving a mean value of 12\,500 K and a median value of
12\,300 K.

Total oxygen abundances (O/H) were calculated using the model-based
code {\sc Hii-Chi-mistry}\footnote{All versions of {\sc HCm} are publicly available on {\url http://www.iaa.csic.es/~epm/HII-Chi-mistry.html}} (hereinafter {\sc HCm;}  version 4.1; \citealt{hcm14,pm19b}).
This code performs a Bayesian-like analysis of several observed emission-line ratios
in comparison with a large grid of photoionization models and calculates a $\chi^2$-weighted mean and standard deviation of the resulting
O/H, N/O, and log $U$. The code makes use of [\oii] $\lambda$3727 \AA,\, [\neiii] $\lambda$3868 \AA,\,
[\oiii] $\lambda$ 4363, $\lambda$5007 \AA,\, [\nii] $\lambda$6584 \AA,\, and [\sii] $\lambda$ 6717+6730 \AA\ reddening-corrected fluxes relative to H$\beta$ and it is consistent with the direct method.
All corrected used emission lines and the calculated oxygen abundances along with all other properties derived in  this work are provided
in electronic format\footnote{At the CDS via anonymous ftp to {\url cdsarc.u-strasbg.fr} (130.79.128.5)
or via \url{http://cdsweb.u-strasbg.fr/cgi-bin/qcat?J/A+A/} }.

We verified that consistency is better than 0.02 dex for 12+log(O/H)
for the 61 objects for which the direct method could be applied (i.e.,
the objects with a direct calculation of $T_e$ and a measurement of
[\oii] $\lambda$ 3727 \AA).  This comparison was made considering
total oxygen abundances from the direct method and the ionization
correction factor (ICF), as calculated in Appendix A.1. In any case,
the ICFs derived for this sample indicate total abundance differences
lower than 0.01 dex.

\begin{figure}
\centering
\includegraphics[width=0.5\textwidth,clip=]{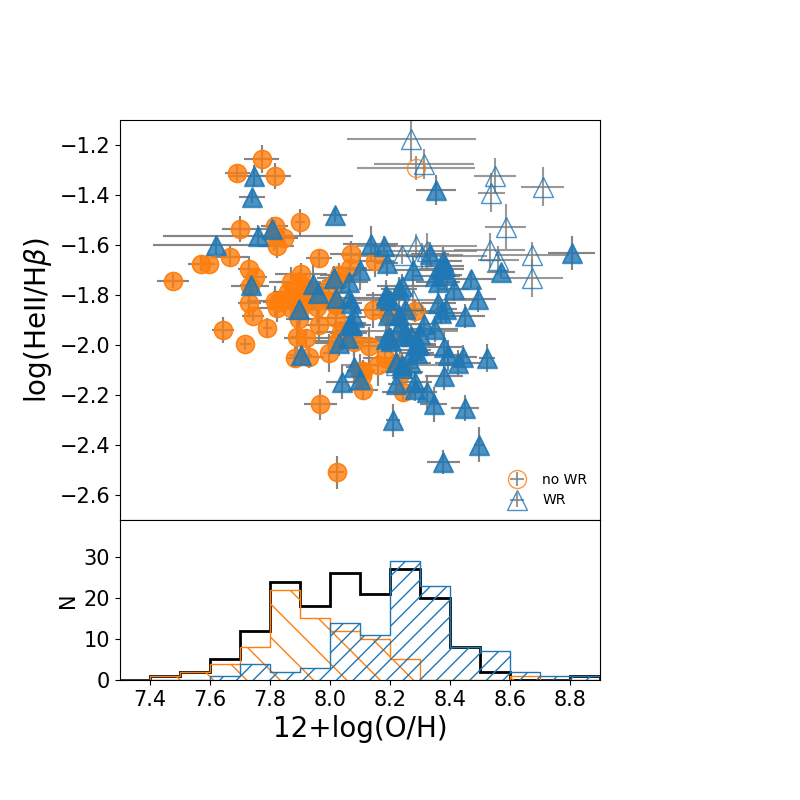}
\caption{{\it Top panel}: Relation between log(\heii/\hb) and the total oxygen abundance calculated 
using {\sc HCm} for objects of the control sample both with the emission-line at [\oiii] $\lambda$ 4363 \AA \ (filled symbols) and without it (empty symbols). Different symbols represent galaxies with a detected bump of WR stars (blue triangles) or without one (orange circles). {\it Bottom panel}: Histogram of the 12+log(O/H). The black line corresponds to the oxygen abundance  distribution for the whole sample, while blue and orange hatched histograms represent the oxygen abundance distributions for WR and nonWR galaxies, respectively.}
\label{oh-he2}
\end{figure}

{\sc HCm} confers the advantage that it can obtain estimations for 12+log(O/H) with uncertainties similar to those
derived from the direct method (i.e., better than 0.03 dex according to \citealt{hcm14}),
when the emission-line ratio [\oiii] 5007/4363 is measured even in absence of the 
 [\oii] $\lambda$3727 \AA\ emission-line.
This is the case for 106 objects in our sample at
redshift $\lesssim$ 0.02, in which the [\oii] lies outside the observed spectral range in SDSS, starting at 3800 \AA.
On the other hand, for those objects without a reliable measurement of the [\oiii] auroral line at $\lambda$4363 \AA \ (i.e., 19 among the selected objects), {\sc HCm} yields O/H
values that are also consistent with the direct method with uncertainties of the order of 0.14 dex according to \cite{hcm14}.
These are represented in the top panel of Fig. \ref{oh-he2}.

In the bottom-panel of Fig. \ref{oh-he2} we present the distribution of the oxygen abundance, 12+log(O/H), calculated from {\sc HCm} for all the \heii \ emitters analyzed in this work.
The  mean value for this distribution is 12+log(O/H) = 8.12 with a $\sigma$ of 0.25 dex (0.27$\times$ Z$_{\odot}$,
taking the solar abundance from \cite{asplund09} as reference).
A consequence of applying a methodology for the derivation of the total oxygen abundance that is consistent with the
direct method for the whole control sample, even when this could not be applied to all objects, 
is that our mean oxygen abundance  is noticeably lower than in the case of \cite{shirazi12} (i.e., 12+log(O/H( = 8.29).
This value underlines the metal-poor nature of this sample which includes six objects that can be considered
as XMPs. 

The mean O/H value for the 106 objects cataloged as WR galaxies is
visibly higher (12+log(O/H)=8.24, with a $\sigma$ of 0.23; see blue histogram in
{\it bottom-panel} of Figure \ref{oh-he2}) than for the 80 objects without the WR bump (12+log(O/H)=7.94,
with a $\sigma$ of 0.19; see orange histogram in
the same panel). We also note that from the six XMPs, only one
shows the WR feature.  All this could be due to the observed and expected
reduced line luminosities of WR stars at low metallicity
\citep[e.g.,][]{schaerer98,ch06,brinchmann08,dori15,bpass}.

In the {\it top panel} of Fig. \ref{oh-he2} we show the relation
between the obtained O/H values and the measured flux of \heii\ at
$\lambda$4686 \AA\ relative to \hb\ in logarithmic units.  The
correlation coefficient ($\rho$) between these two magnitudes for
those objects of the sample whose O/H could be calculated by {\sc HCm}
using the [\oiii] $\lambda$ 4363 \AA\ is -0.36 (-0.09 when all the
objects in the control sample are considered).  The correlation is
detected more clearly for galaxies without the WR bump ($\rho$ =
-0.50) than for those galaxies with it ($\rho$ = -0.11).  This
confirms the trend of higher nebular \heii\ fluxes in lower $Z$
environments in agreement with previous observations
\citep[e.g.,][]{shirazi12,senchyna17}.

\section{Results and Discussion}

\subsection{The softness diagram based on He emission lines}

The relation between \hei\ $\lambda$5876/\heii\ $\lambda$4686 and the
line ratios [\sii]/[\siii] and [\oii]/[\oiii] can constitute a
valuable tool that can be used to trace the slope of the ionizing SED better than other
emission-line ratios such as \heii/\hb. In this case, the SED can be studied from the
first ionization potential of He (i.e.,  24.6 eV) and its shape up to higher energies, at the ionization potential of He$^+$
(i.e., 54 eV), extending the studied range to much higher energies than those of the other involved ions in
the softness diagrams. The other \hei\ prominent emission lines detected in the optical spectrum
($\lambda$4471 \AA, $\lambda$6678 \AA, $\lambda$7065 \AA) lead to
very similar results to those described here, to within the errors, but we focus on the
line at 5876 \AA\ for the sake of higher S/N and simplicity.
For the whole selected control sample, the emission-line ratio \hei/\heii\ shows large dispersion with a mean value in logarithmic units of 0.86 and a standard deviation of 0.25 dex. The difference in this mean value between objects with (0.88) and without (0.84) a WR bump   does not appear to be significant.

\begin{figure*}
\centering
\includegraphics[width=8cm,clip=]{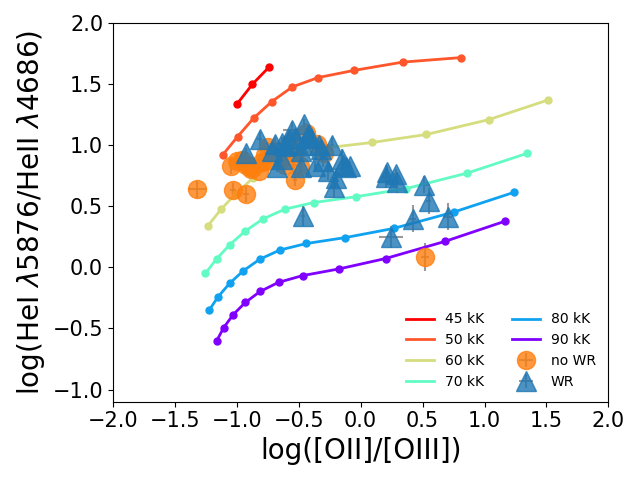}
\includegraphics[width=8cm,clip=]{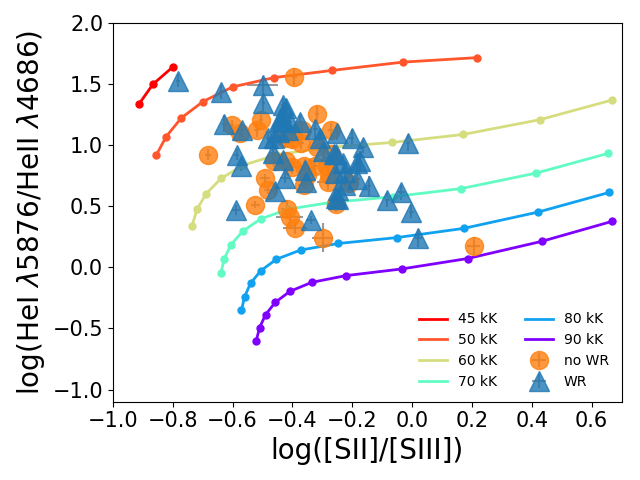}
   \caption{Relation between the emission-line ratio \hei/\heii\ and other optical emission-line ratios, such as [\oii]/[\oiii] 
for 68 objects of the control sample at $z > 0.02$(at left), and [\sii]/[\siii] for another 84 objects at $z < 0.02$ (at right). The different symbols represent objects with (blue triangles) and without (orange circles) an observed WR bump.
The solid lines represent models at different black-body temperatures
of the ionizing source. The points with lower log $U$ are at the upper right of each sequence.}
\label{bb}
\end{figure*}

In Figure \ref{bb} we represent the emission-line ratio \hei/\heii\ in
relation to [\oii]/[\oiii] in 68 objects and in relation to
[\sii]/[\siii] in another 84 galaxies from our sample.  The
emission-line ratios [\oii]/[\oiii] and [\sii]/[\siii] cannot be
used simultaneously in each object of this sample owing to the SDSS spectral range: the
[\oii] $\lambda$3727 line is outside the observed wavelength
range in the SDSS spectra for all galaxies with $z \lesssim $ 0.02, while [\siii]$\lambda$9069 remains out of the observed
spectral range for $z \gtrsim$ 0.02.  In those objects for which we analyzed
the [\sii]/[\siii] emission-line ratio, we considered the theoretical
ratio of 2.44 between I($\lambda$9532 \AA) and I($\lambda$9069 \AA) as
only the latter could be measured in 84 objects.
Therefore, the subsamples represented in Fig. \ref{bb} have different
redshifts. The mean $z$ for the objects in the left panel is 0.056, while this is 0.0058  for objects
in the right panel and these could therefore be more affected by aperture effects.
On the other hand, for another 34 objects in
our sample, neither [\oii] nor [\siii] are available in their spectra
and therefore cannot be represented in any of the two softness
diagrams shown in Fig. \ref{bb}. Alternatively, for such objects, we used the emission line ratio [\sii]/[\oiii]
for our
calculations, which can also be
used to provide a scale for $T_*$ \citep[e.g.,][]{pm19}.

To help us to interpret the diagrams showing [\oii]/[\oiii] ([\sii]/[\siii]) versus \hei/\heii\  , we can use the grid of
photoionization models described in \cite{pm19} which provides us with the emission lines involved in the softness diagrams as a function
of $Z$, $U,$ and $T_*$ for stellar model atmospheres from WM-Basic \citep{wmbasic} in the
range 30-60 kK. 
However, none of the models calculated using these SEDs in this  $T_*$ range are able to predict a measurable flux for the
\heii\ $\lambda$4686 \AA\ line, and so it is necessary to resort to harder SEDs.

To this aim, we built a new grid of photoionization models with the
code {\sc cloudy} v. 17.00 \citep{cloudy} using black-body SEDs as
ionizing sources in the $T_*$ range from 30 to 90 kK in bins of at
most 10 kK.  In addition, we considered values for 12+log(O/H) from
7.1 to 8.9 in bins of 0.3 dex, and for log$U$ from -4.0 to -1.5 in
bins of 0.25 dex. All other ionic species were rescaled to the solar
proportions, with the exception of nitrogen, which was considered as a
primary element for models with 12+log(O/H) $<$ 8.0, and with an extra
secondary production for 12+log(O/H) $>$ 8.0.  We also took into
account a standard galactic dust-to-gas mass ratio
($7.5\times10^{-3}$), a filling factor of 0.1, a constant electron
density of 100 cm$^{-3}$ , and an inner radius of 10 pc, which leads
to a spherical geometry in all cases. The stopping criterion is that
the relative number of free electrons in relation to neutral hydrogen
atoms is not lower than 98\%. The resulting number of models in this
grid is then 924.

In Fig. \ref{bb} we represent some of the derived sequences of models for different values of black-body $T_*$.
 For each set of points at the same $T_*$, the models with lower
 values of $U$ are
shown in the upper right part of the diagrams moving towards higher values of $U$ to the left and downwards.
The models shown in this figure correspond to 12+log(O/H) = 8.0, but no significant differences are obtained when different metallicities are considered.
These diagrams illustrate how 
the model sequences can be used to provide a scale for $T_*$
and, as in the case of the softness diagram based on O and S emission lines, lower values of $\eta'$ parameter correspond to higher values of $T_*$ \cite{pmv09}.
However, given that the sequences of models 
do not have a linear variation and present some additional
dependence on $U$, the relation between them and $T_*$ must be studied using
at least two emission-line ratios to reduce uncertainties.

In this way,  as can be seen, the grid of models cover the observations. All models with $T_*$ $>$ 45 kK predict a certain emission of \heii, and some observed values can only be reproduced with $T_*$ sequences $\gtrsim$ 80 kK, which is much higher than the maximum $T_*$ typical for O and B stars \citep[e.g.,][]{wmbasic}.

\subsection{Using HCm-Teff to derive T$_*$ in \heii \ emitters}

As described in \cite{pm19}, in the case of WM-Basic stellar atmospheres, it is possible to use a grid of models  to perform a Bayesian-like
comparison between the observed emission-line ratios involved in the softness diagrams and the model predictions to calculate estimates of $T_*$ and log $U$.
In the case of our sample of \heii \ emitters, we update the code {\sc HCm-Teff} to version 4.0 to take into account the emission-line ratio
\hei/\heii\ and using the same black-body models described in the above section.

Briefly, the code performs an iterative calculation through the grid of models interpolated at the metallicity derived for each set of observations
and computes the $\chi^2$-weighted mean and standard deviation for both $T_*$ and log $U$ given by each model.
The $\chi^2$ weights are calculated as the quadratic difference between the
observed and predicted   emission-line ratios:
[\oii]/[\oiii], \hei/\heii, and [\sii]/[\siii] when available. In addition, the code can use [\sii]/[\oiii] in cases where [\oii] and/or [\siii] emission lines are
not available.
As a test, we verified that when we use the same emission lines predicted by each model  as an input, we recover the values
of $T_*$ with an uncertainty lower than 500 K and the values of log$U$ with an uncertainty lower than 0.05 dex across the whole range of values, which are very close to the uncertainties
found in \cite{pm19} using other stellar atmospheres and without He lines.
Table \ref{test} lists the mean offsets and the standard deviation of the residuals between the values of $T_*$ and log $U$
derived from {\sc HCm-Teff} as a function of different emission-line ratios used as input from the grid of models.

\begin{table*}
\begin{minipage}{180mm}
\begin{center}
\caption{Mean offsets and standard deviation of the residuals of the resulting properties derived
by {\sc HCm-Teff} when the model emission lines are used as input as a function of
the used emission-line ratios.}
\begin{tabular}{lcccccccc}
\hline
\hline
    Used line ratios & \multicolumn{2}{c}{$T_*$ (kK)\footnote{From models using black-body SEDs}} & \multicolumn{2}{c}{log $U^a$} & \multicolumn{2}{c}{log $f_{abs}$\footnote{From models using BPASS v.2.1 for $Z_*$ = $Z_g$, $x$ = -1.35 and M$_{up}$ = 300 M$_{\odot}$ SEDs and density-bounded geometry}} & \multicolumn{2}{c}{log $U^b$} \\
   & Mean $\Delta$ & $\sigma$ res. & Mean $\Delta$ & $\sigma$ res. & Mean $\Delta$ & $\sigma$ res. & Mean $\Delta$ & $\sigma$ res.  \\ 
\hline
[\oii]/[\oiii], [\sii]/[\siii], \hei/\heii &  +0.8 &  2.1   & +0.01 & 0.10 & +0.11  &  0.34  &  -0.10 & 0.53  \\
~[\oii]/[\oiii], \hei/\heii &  +0.8 &  2.1   & +0.01 & 0.10 & -0.12  &  0.27  &  +0.14 & 0.47  \\
~[\sii]/[\siii], \hei/\heii &  +0.6 &  2.0   & +0.03 & 0.13 & -0.19  &  0.35  &  +0.10 & 0.62  \\
~[\sii]/[\oiii], \hei/\heii  &  -0.3 &  5.1   & +0.04 & 0.51 & -0.07  &  0.27  &  -0.10 & 0.59  \\
~\hei/\heii  &  -0.5 &  6.8   & +0.04 & 0.72 & -0.12  &  0.35  &  -0.25. & 0.63  \\
\hline

\label{test}
\end{tabular}
\end{center}
\end{minipage}
\end{table*}

   \begin{figure*}
   \centering
\includegraphics[width=8cm,clip=]{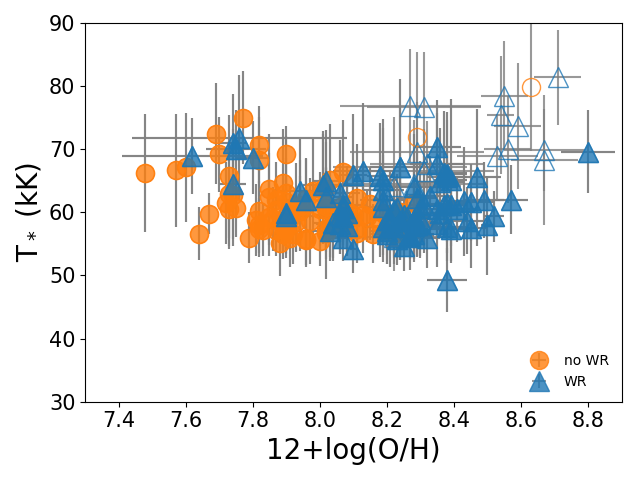}
\includegraphics[width=8cm,clip=]{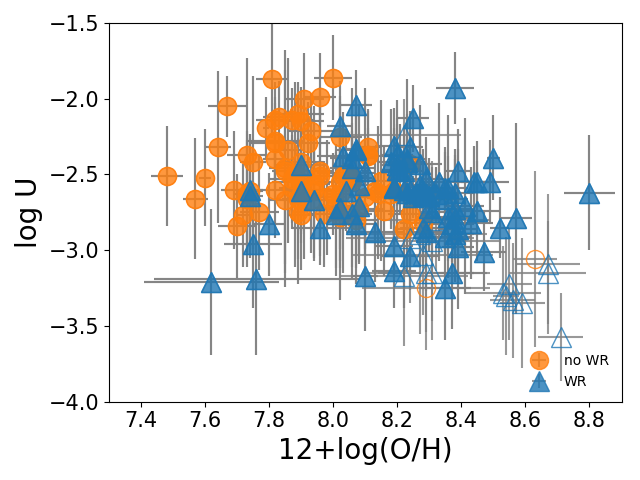}

      \caption{Relation between total oxygen abundance and the values obtained
from {\sc HCm-TEff} for our sample of \heii \ emitters assuming a black-body temperature: with $T_*$ (left) and with log $U$ (right). Different symbols represent objects with (blue triangles) or without (orange circles) the WR bump.
Filled symbols represent objects for which O/H was calculated using the [\oiii] $\lambda$ 4363 \AA, while empty symbols represent objects for which O/H was not calculated using this line.}
         \label{Teff-U}
   \end{figure*}

According to the results from {\sc HCm-Teff}, the mean $T_*$ for the 186 \heii \ emitters is
61\,900 K, with a standard deviation of 5\,500 K, and the obtained range goes
from $\sim$ 50 kK to around 80 kK. The mean value of $T_*$ is slightly higher in the case of objects with a WR bump (62\,400 K)
than for objects without one (61\,200 K).
On the other hand, we did not find significant differences between the mean $T_*$ for those objects at the higher redshift 
for which we used the  [\oii]/[\oiii] ratio (61,500 K, with $\sigma$ = 6,000) and  those at the lowest redshift for which we
used the [\sii]/[\siii] emission line-ratio (63,000 K, with $\sigma$ = 5,000 K).  
The left panel of Figure \ref{Teff-U} shows the relation between the obtained $T_*$ and total oxygen abundance
for the 186 objects of the sample.
A very weak correlation is observed between them, with $\rho$ = -0.15
($\rho$ = -0.21 considering only those objects for which O/H was calculated using the [\oiii] $\lambda$ 4363 \AA\ line), meaning that objects with lower $Z$ appear to have 
higher $T_*$ on average. As a side note, the mean $T_*$ derived for the
XMPs is slightly higher (64\,200 K) than the average value  for the sample, but these differences are not significant as they are lower
than the obtained typical errors of around 6\,000 K.

Regarding $U$, the mean value for the whole sample is log $U$ = -2.58,
with a standard deviation of 0.26 dex, and the entire range covers
-3.25 to -1.86.  As in the previous case for $T_*$, no significant
difference is found when the [\oii]/[\oiii] is used (mean log $U$ of
-2.60) as opposed to when [\sii]/[\siii] (-2.66) is used for those
objects at the lowest redshift.  In the right panel of Figure
\ref{Teff-U} we present the relationship between the obtained $U$
values and metallicity; as in the case of $T_*$, a mild correlation
can be seen ($\rho$ = -0.41).  Galaxies with lower metallicity have on
average higher excitations in agreement with the result for SF
galaxies and \hii\ regions found by \cite{hcm14}.  This dependence
between metallicity and $U$ can explain the lower mean log $U$ value
found for WR objects (-2.66) in comparison to objects without a WR
bump (-2.46), because the latter present lower O/H on average in this
sample, as discussed above. However, as in the case of $T_*$, as the
typical error obtained for log$U$ is around 0.3 dex, this result
cannot be taken as significant.

\subsection{Compatibility with cluster model atmospheres}

Considering that black-body SEDs are not realistic and they were  only used in order to
obtain an absolute scale of which is not possible for the atmospheres
  of massive stars, we may wonder to what extent it is possible to
  reproduce the very high values of $T_*$ found for the studied sample
  of \heii \ emitters using increasingly realistic SEDs. 

To this aim, we produced additional grids of photoionization models
using cluster model atmospheres from {\sc BPASS} v.2.1 \citep{bpass} under
different conditions: We considered ionizing SED clusters
assuming binarity and instantaneous  star formation bursts with ages from 1 to
  10 Myr in steps of 1 Myr, two values for the slope of the IMF $x$ = -1.00 and -1.35, upper mass limits of
  100 and 300 M$_{\odot}$, and two metallicities ($Z_*$ = 0.0022,
  approximately equivalent to 12+log(O/H) = 8.0; and nearly metal-free stars at $Z_*$ = 10$^{-5}$).
For the gas we considered a mean oxygen abundance of 12+log(O/H) = 8.0, with the remaining chemical species 
scaled to the solar proportions according to \cite{asplund09}, and log$U$ = -2.5,
which is close to the average values found in the sample of \heii \ emitters.  We also assumed 
in the models all the other input  conditions described in the above sections.

%
   \begin{figure*}
   \centering
\includegraphics[width=8cm,clip=]{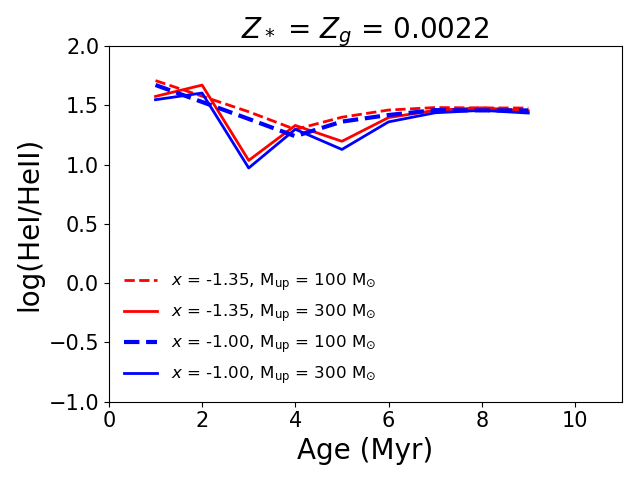}
\includegraphics[width=8cm,clip=]{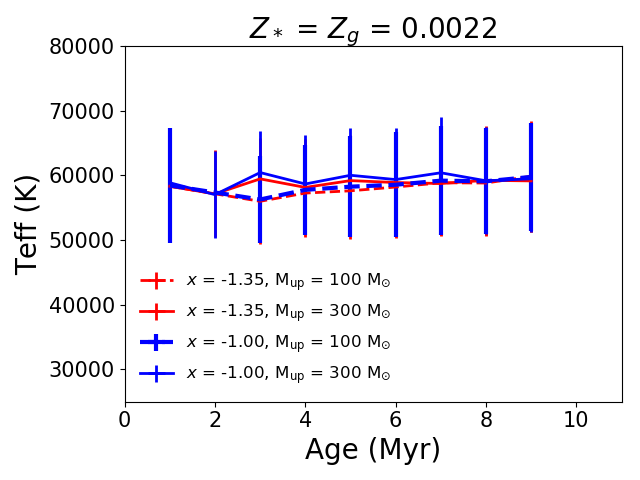}
\includegraphics[width=8cm,clip=]{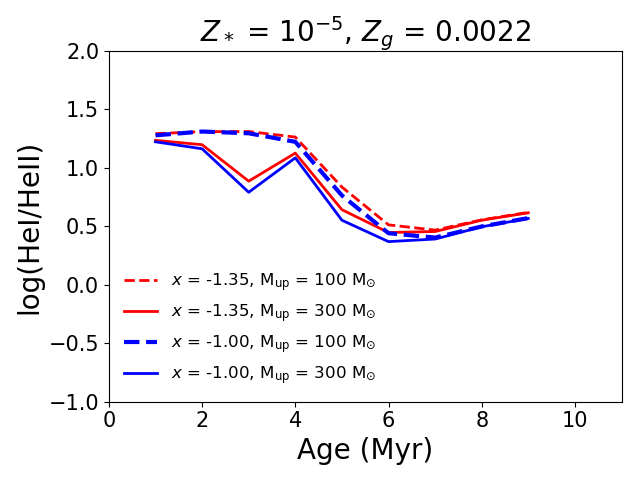}
\includegraphics[width=8cm,clip=]{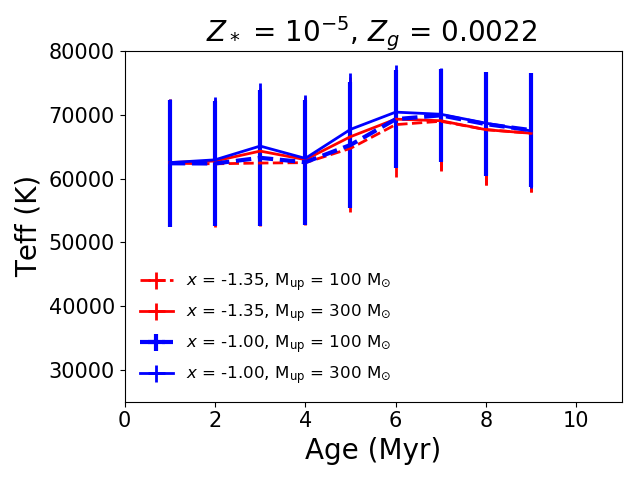}

      \caption{Evolution of the emission-line ratio \hei/\heii\ (left column) and $T_*$ (right column) from evolutionary synthesis atmospheres from {\sc BPASS} v. 2.1 for different IMF and upper mass limits and metallicities as predicted from {\sc Cloudy} models and {\sc HCm-Teff}. The $T_*$ values were derived with {\sc HCm-Teff} and vertical error bars show the uncertainty of these values as derived by the code. Panels in the upper row show predictions assuming the same $Z$ both for the stars and the gas (i.e., $Z_*$ = $Z_g$ = 0.0022), while the panels of the 
lower row show predictions assuming nearly metal-free stars ($Z_*$ = 10$^{-5}$ and $Z_g$ = 0.0022).}
         \label{bpass}
   \end{figure*}

The upper-left panel of Fig. \ref{bpass} shows the predictions from the models for the \hei/\heii\ emission-line ratio as a function of the age
of the burst for the different considered IMF conditions in the models and assuming the same metallicity both for stars and gas ($Z_*$ = $Z_g$ = 0.0022). 
As can be seen,
all models reach the minimum value for this ratio at an age of between 3 and 6 Myr, which is when  the WR phase is expected to take place for massive stars, confirming  the important role of this stage in the hardening of the
incident radiation. 
As also expected, the \hei/\heii\ emission-line ratio is lower when we assume a higher upper limit for the stellar mass,
and above all a flatter slope for the IMF; the minimum value for log(\hei/\heii)  is 0.93,
for $x$ = -1.0 and M$_{up}$ = 300 M$_{\odot}$. However, this value is
larger than the average log(\hei/\heii) measured in our sample of \heii-emitters of 0.86, being of 0.24 at 2.5$\cdot \sigma$. 

The same conclusions are reached when we use {\sc HCm-Teff} to transform the predicted emission lines from the models using these
stellar cluster atmospheres into a scale of $T_*$ from the black-body models, as described in the previous section. 
These results are shown in the upper right
panel of Figure \ref{bpass}. As in the case of the emission-line ratio \hei/\heii, but with an opposite sign, the maximum value for $T_*$ is reached during
the WR phase between 3 and 6 Myr for most models. The maximum $T_*$ is reached assuming a flat top-heavy IMF with an upper mass limit of 300 M${_\odot}$. However, this maximum $T_*$ is only
59\,300 K, which  is lower than the mean value found in the observed sample of \heii \ emitters,
namely around 62\,000 K.
We also verified that similar values are reached when we consider, in these models, values for $Z_g$ according to  the whole  derived distribution of 12+log(O/H) for our sample of \heii\ emitters.

Another factor, apart from the IMF conditions, that can harden the
ionizing SED is the metallicity of the stars. It has been suggested
that nearly metal-free stars can have
a much more energetic flux for frequencies higher than 4 Ryd, which are able to ionize He$^+$ \citep{schaerer03,ts00,ku19}, and so we also produced models with stars at the lowest available metallicity in {\sc BPASS} v. 2.1, which
is of $Z_*$ = 10$^{-5}$ and assuming the same conditions as those described above for the grid with the same metallicity, both for stars and gas. 
The predictions from these models 
for log(\hei/\heii) are represented in the  lower left panel of Figure \ref{bpass}.
As can be seen, in this case the models reach  a minimum log(\hei/\heii) value (0.38  at an age of 6 Myr for models with $x$ = -1.00 and an upper mass limit of 300 M$_{\odot}$) that
is much lower than the mean of the observed distribution of log(\hei/\heii)  in our sample of \heii \ emitters.
The rest of the models with $Z_*$ =  10$^{-5}$,  assuming different IMF conditions, also reach a
minimum value for \hei/\heii\ that is lower than the mean value observed for the sample of \heii -emitters.
The  lower right panel of Figure \ref{bpass} shows the evolution of the
equivalent $T_*$ derived from these same models with nearly metal-free stars using the predicted emission lines
required by {\sc HCm-Teff}. In an equivalent way to the previous case, the highest value for $T_*$ is reached at an
age of around 6 Myr for models with an IMF slope of $x$ = -1.00 and an upper mass limit of 300 M$_{\odot}$.
This highest  derived value (66\,500 K) is above the mean value derived 
in our sample of selected \heii -emitters.

Figure \ref{fabs} shows the softness diagrams which portray the \hei/\heii\ emission-line ratio
against [\oii]/[\oiii] (for 68 objects of the sample at $z > 0.02$) and against [\sii]/[\siii] (for another 84 objects at $z < 0.02$), as compared
with our {\sc Cloudy} models both for $Z_* = Z_g = 0.0022$ and for $Z_*$ =10$^{-5}$, assuming different values for log $U,$ the ages of the instantaneous bursts at the predicted minimum value for log(\hei/\heii), and a radiation-bounded geometry (absorption factor, $f_{abs}$ = 1).
As can be seen, in principle one could estimate the metallicity of the ionizing stars using this method if the age of the burst is known in a part of the sample.
Nevertheless, although models assuming nearly metal-free stars from {\sc BPASS} are
able to closely reproduce
the average distributions observed both for \hei/\heii\ and $T_*$, 
these models cannot reach the observed and derived ranges beyond 1$\cdot \sigma$ in the corresponding distributions
(e.g., log(\hei/\heii) reaches up to 0.36 at 2$\cdot sigma$ below the mean, and $T_*$ can reach up to 72\,900 K at 2$\cdot sigma$ above the derived mean $T_*$ value).
Thus, the observed and derived ranges 
both for \hei/\heii\ and $T_*$ in our sample of \heii \ emitters cannot be entirely covered only by invoking variations
of the assumed IMF, or the age and metallicity of the ionizing stars, and therefore additional
assumptions need to be made in order to fully understand the behavior of our sample in the softness diagrams.

\subsection{Using the softness diagrams to derive the photon escape fraction}

A further scenario to be explored in order to explain the mean value and corresponding dispersion, as well as the range of the distributions of the \hei/\heii\ ratio and the derived $T_*$, includes the effect of photon leaking. A natural way to simulate this effect is a density-bounded geometry, which implies that the low-excitation region of the gaseous nebulae cannot be completed and therefore the high-excitation lines would have more weight in the integrated spectrum of the gas.
This effect has been suggested by several authors \citep[e.g.,][]{nakajima14,Paalvast2018} who proposed the
[\oiii]/[\oii] emission-line ratio (dubbed O32) as a tracer for photon leaking when it reaches very high values.
Furthermore, \cite{izotov16} and \cite{izotov18}   used measurements of the Lyc leaking to confirm that certain SF galaxies 
do not retain all their ionizing photons and, at the same time, present very high values of O32.
Nevertheless, considering that O32 also depends on other functional parameters of the gas,
such as metallicity, $U,$ and $T_*$, the use of the different versions of the softness diagram provides a complementary diagnostic of this effect, which can be fully exploited when the \heii\ emission line is observed.

To explore this possibility we produced additional grids of photoionization models in a very
similar way to those described in previous sections. In this case, we used the SEDs from {\sc BPASS} v.2.1 cluster models,
assuming binarity, instantaneous star formation, and an IMF with a slope of $x$ = -1.35 and an upper mass limit of 300 M$_{\odot}$.
Regarding the age and metallicity of the stars, we built a grid for the lowest available $Z_*$ at 10$^{-5}$, 
with an age of 6 Myr, in agreement with the age for which the minimum \hei/\heii\ is reached for this metallicity.
In addition, we built a grid of models for which $Z_*$ = $Z_g$, covering the range of possible values of $Z_g$, with an age of 4 Myr.
For the gas we assumed different O/H values from 12+log(O/H) = 7.1 to 8.9 in bins of 0.3 dex, scaling
the remaining elements in the same way as described in Section 3.2.
We also considered  values for log $U$ from -4.0 to -1.5 in bins of 0.25 dex. The remaining gas conditions were set equal to those of the other model grids described above. 
This implies  77 models for each grid so far.
>From these, we recalculated the models, now changing  the stopping criterion in such a way that the
number of absorbed photons ionizing H is a fraction of the corresponding quantity in radiation-bounded models. For each value of metallicity $Z,$ we calculated models
at an $f_{abs}$ of 0.5, 0.2, 0.1, 0.05, 0.02, and 0.01. The resulting total number of models built is 539 for each grid.  

   \begin{figure*}
   \centering
\includegraphics[width=8cm,clip=]{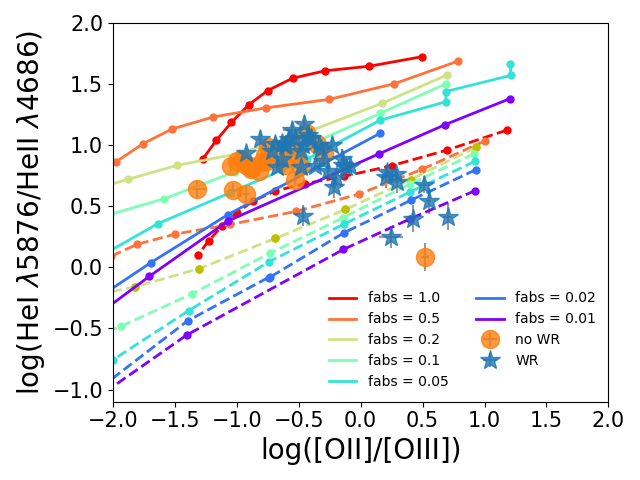}
\includegraphics[width=8cm,clip=]{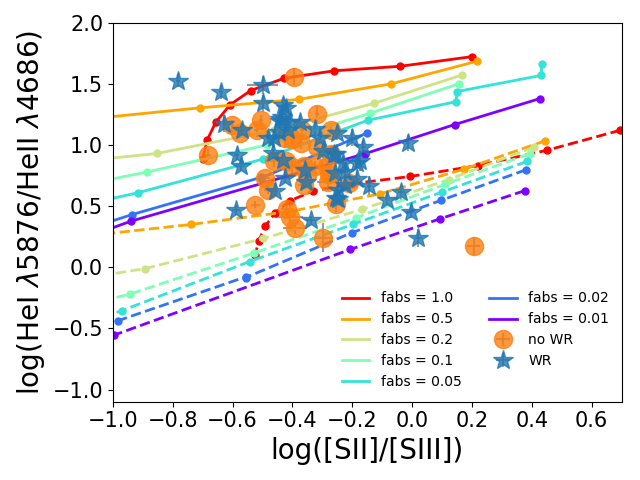}

   \caption{Relation between the emission-line ratio \hei/\heii\ and the line ratios [\oii]/[\oiii]
for 68 objects of the control sample at $z > 0.02$
(at left), and [\sii]/[\siii] for another 84 objects at $z < 0.02$ (at right). Different symbols represent objects with (blue triangles) and without (orange circles) an observed WR bump. The lines represent predictions from models
using {\sc BPASS} 2.1 SEDs with an IMF with $x$ = -1.35 and an upper
mass limit of 300 M$_{\odot}$ considering different fractions of
absorbed ionizing photons. Solid lines stand for models at 4 Myr with
$Z_*$ = $Z_g$ = 0.0022, while dashed lines represent models at 6 Myr
with $Z_*$ = 10$^{-5}$ and $Z_g$ = 0.0022. Log U decreases towards the top right corner of each panel.}
              \label{fabs}%
    \end{figure*}

In Figure \ref{fabs} we repeat what it is shown in Figure \ref{bb}, but changing the sequences of models
for different values of $f_{abs}$ and $Z_*$. As can be seen, the different sequences mostly cover the distribution of the observed emission-line
ratios in both panels. As in the case of a black-body grid, U decreases towards the upper
right corner of each panel and increases towards the lower left corner.
The models assuming $Z_*$ = $Z_g$ predict very high values of
\hei/\heii\ for $f_{abs}$ = 1, and they can cover an important
fraction of the distribution of points for lower values of
$f_{abs}$. On the other hand, models assuming nearly metal-free stars and $f_{abs}$ = 1 can reach much lower \hei/\heii\  in the observed galaxies, while for encompassing the still uncovered part of the distribution of galaxies, lower values of $f_{abs}$ have to be assumed. In all cases, these sequences are also dependent on the age that we assume for the stars and, to a lesser extent, on the properties of the IMF.  

   \begin{figure*}
   \centering
\includegraphics[width=8cm,clip=]{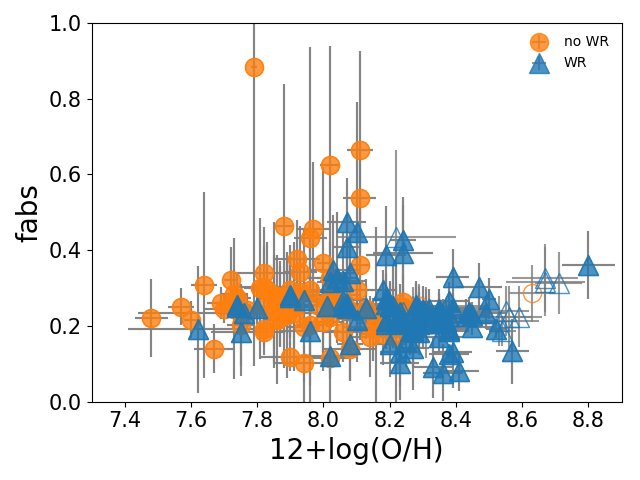}
\includegraphics[width=8cm,clip=]{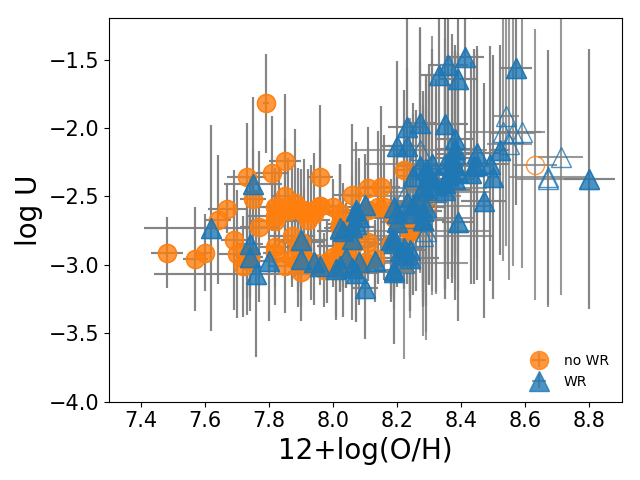}

      \caption{Relation between the total oxygen abundance for the sample of \heii \ emitters and the fraction of absorbed photons (left panel) and the ionization parameter (right panel) as derived from {\sc HCm-Teff}.
Different symbols represent objects with (blue triangles) or without (orange circles) the WR bump in their spectra.
Filled symbols represent objects for which O/H was calculated using the [\oiii] $\lambda$ 4363 \AA, while empty symbols represent objects for which O/H was not calculated using this line.}
         \label{fabs-U}
   \end{figure*}

We adapted the {\sc HCm-Teff} code to provide solutions from the grids of models for both $f_{abs}$ and log $U$ for the sample galaxies.
In this case, we only considered in the code the models described above that assume identical metallicity for the gas and stars, $Z_*$ = $Z_g$, seeking to ensure that the $f_{abs}$ solutions derived   
match within the larger area of the softness diagrams covered by our sample of \heii \ emitters.
The procedure is identical to that described in Section 3.3, but changing the models so that the final product calculated by the code is now $f_{abs}$ instead of $T_*$.
We verified that when we use the emission lines predicted by each model  as input for {\sc HCm-Teff}  we recover their corresponding $f_{abs}$ and log$U$ values with uncertainties better than 0.2 and 0.5 dex, respectively.
The mean offsets and standard deviation of the residuals as a function of the emission-line ratios used as input are listed in Table \ref{test}.

>From this analysis, all objects could have in principle photon leaking to some extent, with a mean value for $f_{abs}$ of 0.26 and a standard deviation of 0.10;
the obtained range goes from almost 0.9 down to $\approx$ 0.1.
As in the case of $T_*$, no significative difference is found between the objects at higher redshift for which we used [\oii]/[\oiii] ($f_{abs}$ = 0.24)
and the objects at lower redshift for which we used [\sii]/[\siii] ($f_{abs}$ = 0.29).
The mean absorption fraction is slightly lower in the case of galaxies with a detected WR bump ( $f_{abs}$ = 0.24) than in those without it ($f_{abs}$ = 0.28), but this difference cannot be taken as significant given the typical error for $f_{abs}$ of 0.1.

The left panel of Figure \ref{fabs-U} shows the obtained values of $f_{abs}$ as a function of O/H
for the whole sample.
As in the case of $T_*$, the data appear to show a slight correlation, $\rho$ = -0.20 
(-0.27 when we only consider objects for which O/H was calculated using [\oiii] $\lambda$ 4363 \AA), suggesting that objects of lower $Z$ tend to have higher absorption fractions; although this trend appears to concern mostly WR objects for which $\rho$ = -0.29, while for  objects without the WR bump we find $\rho$ = -0.05.

Regarding log $U$, the mean value obtained for the sample from this grid is -2.50 with a
standard deviation of 0.40 dex. However, given that the mean error obtained from
this method for log $U$ is 0.5 dex, it is difficult to draw any significant conclusions
from the $U$ values from the density-bounded grid, as the $U$ values predicted by the models dramatically increase for low values of $f_{abs}$.
Nonetheless, if we restrict our analysis to those objects with an error for log$U$ of lower than 0.3 dex (only 32 objects in the sample), the mean log $U$ value is signifcantly lower (-2.85) with no significant differences between objects with or without the WR bump.

The right panel of  Figure \ref{fabs-U} shows the relation between the obtained log $U$ from {\sc HCm-Teff} under the assumption of density-bounded geometries and O/H for those objects with the [\oiii] $\lambda$4363 \AA\ line. 
In this case we see a positive correlation between the two properties ($\rho$ = 0.66), contrary to the 
anti-correlation between $U$ and $Z$ observed when we derive $T_*$. However, if we restrict our
analysis to those objects with an error in log $U$ of lower than 0.3 dex, the correlation coefficient changes its sign ($\rho$ = -0.20),  in better agreement with previous results.

Altogether, though the derivation of log $U$ from this density-bounded model grid remains somewhat uncertain, these models seem to reproduce the overall trend of this parameter covering a sizeable part of the softness diagrams, especially when invoking different absorption fractions of ionizing photons $f_abs$. Nevertheless, the values obtained for $f_abs$ strongly depend on the assumptions made for the IMF, age, or metallicity of the ionizing stars. In principle,  the derived $f_{abs}$ values are expected to increase for those models assuming harder SEDs (e.g.,  lower $Z_*$; more top-heavy IMF) or decrease for the softer SEDs (e.g.,  ages before the WR bump; lower upper mass limit of the IMF).

\begin{table}
\begin{minipage}{90mm}
\begin{center}
\caption{Mean and dispersion of observed and derived properties in the studied sample of \heii \ emitters.}
\begin{tabular}{lccc}
\hline
\hline
   &  Total  &  no WR   &  WR \\
\hline
Number &   186  &  80  &   106  \\
log([\oii]/[\oiii]  &  -0.44 (0.43)  &  -0.70 (0.33)   &  -0.27 (0.40)   \\
log ([\sii]/[\siii])  &  -0.37 (0.17)  &  -0.38 (0.15)  &  -0.34 (0.18)  \\
log(\hei/\heii)  &  0.86 (0.25)   &   0.84 (0.24) & 0.88 (0.26)   \\
12+log(O/H)  &   8.11 (0.26)  & 7.94 (0.20)   &    8.24 (0.23) \\
T$_*$ (kK)\footnote{From models using black-body SEDs}  &   61.9 (5.1)  &    61.2 (5.0)  &  62.4 (5.2)  \\
log $U$$^a$  &    -2.58 (0.28) &   -2.46 (0.25)  &  -2.66 (0.28)  \\
$f_{abs}$\footnote{From models using BPASS v.2.1 for $Z_*$ = $Z_g$, $x$ = -1.35 and M$_{up}$ = 300 M$_{\odot}$ SEDs and density-bounded geometry}  &   0.26 (0.10) &    0.28 (0.12)  &  0.24 (0.08) \\
\hline

\label{mean}
\end{tabular}
\end{center}
\end{minipage}
\end{table}

\section{Summary and conclusions}

In this work, we present a study of 186 SF galaxies with prominent nebular \heii\ $\lambda$4686 \AA\ taken
from SDSS-DR7, a sample previously studied by \cite{shirazi12}. We investigated their behavior in the so-called softness diagram, which is useful to
derive the scale of equivalent effective temperature in combination with the ionization parameter.
The inclusion of the emission-line ratio \hei/\heii\ in this diagram allows us to more accurately explore
the hard ionizing sources in \heii \ emitters as it better traces the high-energy range in their SEDs.

Our re-analysis of this sample allows us to derive accurate total oxygen abundances by means of the code {\sc HCm}, taking advantage of the measured electron
temperature in 167 objects of the sample. We were also able to measure for the first time the [\siii]$\lambda$9069 line
in 84 objects of the sample at $z \lesssim$ 0.02. This allows us to use [\sii]/[\siii]
in the softness diagram of these objects, while [\oii]/[\oiii] could be used for the  objects at $z \gtrsim$ 0.02 where [\siii] is not covered in
the observed spectral range.

The new derived metallicities underline the metal-poor nature of these galaxies (mean 12+log(O/H) = 8.11), but 
significant differences are found between the galaxies with a detected WR bump (12+log(O/H) = 8.24) and those without (12+log(O/H) = 7.94). The relation between the metallicity
and the relative flux of \heii \ to \hb, which is higher on average for the most metal-poor galaxies, is more clearly observed
for galaxies without the WR bump, confirming the result already obtained by \cite{shirazi12}.

Our analysis indicates that the observed softness diagrams involving He lines of these objects can be reproduced using photoionization models calculated
with black-body SEDs with temperatures in the range 45-90 kK, which is far wider than the range explored by individual massive star atmosphere models.
The application of the code {\sc HCm-Teff} to perform a Bayesian-like analysis in the space of these models covering a
wide range of metallicities $Z$, equivalent effective temperatures $T_*$, and log $U$, indicates that the mean $T_*$ for the galaxy sample is higher than 60\,000 K. The mean $T_*$ and
its corresponding standard deviation  as a function of the presence or absence of a WR bump in these galaxies is shown in Table \ref{mean}.
Given that no significany differences are found between the results for the galaxies with the WR bump and those without, there is no clear evidence that WR stars directly constrain the position of \heii -emitting galaxies in the softness diagram. 

When we try to reproduce the results obtained above using more realistic model cluster atmospheres from {\sc BPASS} v. 2.1 \citep{bpass},
we find that the mean of the observed distributions for \hei/\heii\ and for $T_*$ can only be reached when we assume the lowest available metallicities for the stars (i.e., $Z$ = 10$^{-5}$).
This finding gives support to previous results on the total \heii\ budget found in some extremely metal-poor galaxies as compared with evolutionary models
\citep{k18,st19}. 
Changing other properties of the ionizing clusters in the models, such as the age, the IMF
 slope, or the upper mass limit, does not introduce significant differences in the resulting $T_*$
(See also \citealt{st19}).

Alternatively, the inclusion of different density-bounded geometries in the models, even assuming that the metallicity of the stars is the same as that measured in the gas, allows us to cover the observed mean values and nearly the entire range for \hei/\heii\ and $T_*$. In this case, our models indicate that most objects would present some degree of photon leaking, resulting in a mean absorption fraction for our galaxy sample of $f_{abs}$ = 26\%. Nevertheless, the concrete values of the absorption fraction should vary depending on the $Z$ assumed for the ionizing stars and also on the different ages and IMF properties assumed in the synthetic cluster models.

These findings suggest that there is no need to invoke other additional ionizing sources (e.g., shocks, X-ray binaries, AGNs), in agreement with recent results from cosmological hydrodynamical simulations \citep[e.g.,][]{barrow2020}, and imply a very similar solution to the discrepancy between Zanstra temperatures and observations of planetary nebulae showing \heii\ emission
\citep{tilenda94}.
As a consequence, both factors (nearly metal-free stars and density-bounded geometry) appear to be fundamental and have a non-negligible effect on 
the observed behavior of \heii\ emitters in the softness diagram.
Moreover, the co-existence of these factors could be supported by a causal link between
photon leaking in starburst galaxies and moderate or extreme outflows \citep[e.g.,][]{Weiner2009,Chisholm2017, Berg2019}; if these outflows are mainly driven by the star formation processes, free paths around hot star clusters can be created facilitating the leakage of ionizing photons.
In any case, further studies of large, statistically significant
samples of \heii \ emitters involving more emission-line ratios are
needed to shed more light on the role played by these fundamental
factors (i.e.
photon leaking versus hardness of the ionizing source) in the interplay between stars and gas in these objects.

\begin{acknowledgements}
We thank an anonymous referee for his/her constructive comments and suggestions that have helped us to improve this manuscript.
We acknowledge financial support from the State Agency for Research of the Spanish MCIU through the "Center of Excellence Severo Ochoa" award to the Instituto de Astrof\'\i sica de Andaluc\'\i a  (SEV-2017-0709).
This work has been partly funded by the Spanish  Ministerio de Econom\'ia y Competitividad project "Estallidos6" AYA2016-79724-C4 and Junta de Andaluc\'\i a excellence grant EXC/2011 FQM-7058. RGB acknowledges financial support from the Spanish Ministerio de Econom\'ia y Competitividad, through 
projects AYA2016-77846-P and AYA2014- 57490-P.
EPM also acknowledges  the assistance from his guide dog Rocko without whose daily help this work would have been much more difficult.

\end{acknowledgements}

%
   \bibliographystyle{aa} 
   \bibliography{HeII.bib} 
%

\begin{appendix}
\section{Impact on ionization correction factor}

The measurement of \heii\ emission lines in gaseous nebulae implies the existence
of a very hard ionizing radiation source able to create highly ionized species. In the case of oxygen, the most abundant
and representative metal species in the gas, this can lead to a certain amount of ions being more highly ionized than those species whose 
abundances are estimated from the lines in the optical spectrum, namely O$^+$ and O$^{2+}$. This correction factor can be expressed in the following terms:

\begin{equation}
\frac{\rm O}{\rm H} = {\rm ICF}({\rm O}^+ + {\rm O}^{2+}) \times \frac{{\rm O}^++{\rm O}^{2+}}{{\rm H}^+}
.\end{equation}

\begin{figure*}
   \centering
\includegraphics[width=8cm,clip=]{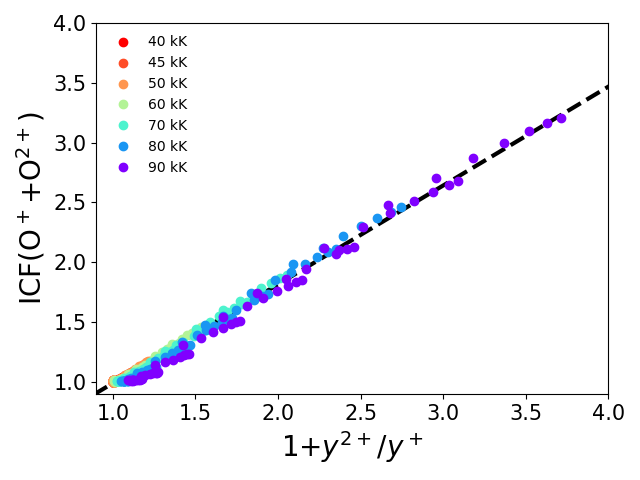}
\includegraphics[width=8cm,clip=]{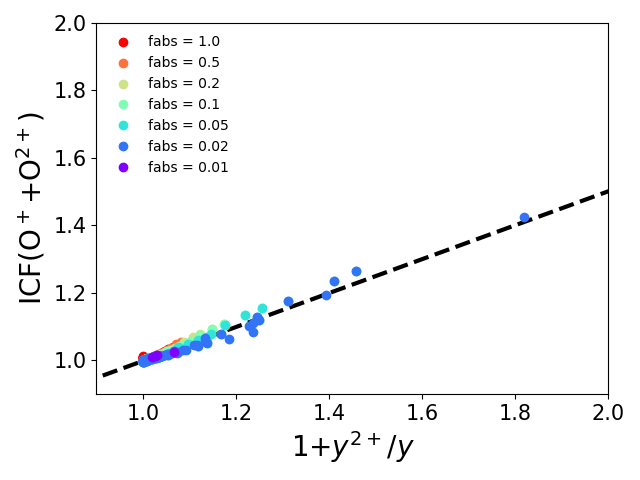}

      \caption{Relation between the ionization correction factor for $y^+$ and that for O$^+$+O$^{2+}$ as predicted from models and 
the respective linear fit. (Left) Models for black bodies of different temperatures. (right) Models from BPASS v.2.1 at an age of 4 Myr assuming $Z_g$ = $Z_g$, with an IMF of slope $x$ = -1.35 and an upper stellar mass of 300 M$_{\odot}$,  calculated for different absorption factors for the number of ionizing photons.}

         \label{icf}
   \end{figure*}

In Figure \ref{icf} we show the relation between this ICF for oxygen and the ICF obtained for $y^+$ (i.e., the relative abundance of He$^+$), which can be easily calculated when the \heii\ emission is detected, using:

\begin{equation}
{\rm ICF}(y^+) = 1 + \frac{y^{2+}}{y^+}
.\end{equation}

As can be seen, the relation between both ICFs is linear in all cases, but the slope of this relation depends mostly on the shape of the ionizing SED. In the left panel of Fig. A1 no
variation is detected as a function of the effective temperature of the black-body used in the models. 
This slope is:

\begin{equation}
{\rm ICF}({\rm O}^++{\rm O}^{2+}) = (0.827 \pm 0.003) \times \left(1+\frac{y^{2+}}{y^+}\right) + (0.162 \pm 0.004)
.\end{equation}

In contrast, this slope is much higher than that obtained using BPASS v.2.1 models for an age of 4 Myr, an IMF slope of 1.35, and an upper stellar mass limit of 300 M$_{\odot}$, assuming $Z_*$ = $Z_g$. In this case, the obtained expression is, 

\begin{equation}
{\rm ICF}({\rm O}^++{\rm O}^{2+}) = (0.503 \pm 0.004) \times \left(1+ \frac{y^{2+}}{y^+}\right) + (0.494 \pm 0.004)
.\end{equation}

Nevertheless, this slope does not change for different fractions of absorbed ionizing photons in the models, and so we can conclude that the main factor
contributing to the calculation of this ICF is not the density-bounded geometry of the gas.
Therefore, the ICFs depend mostly on the SED used in the models, but do not change as a function of the different density-bounded geometry assumptions.

\end{appendix}

\end{document}